\documentclass[]{aa}
\usepackage{graphicx}
\usepackage{epsfig}
\usepackage{longtable}
\usepackage{lscape}

\begin{document}

\title{The Globular Cluster System of NGC\,5846 Revisited: Colours, Sizes and X-Ray Counterparts \thanks{Tables 3 and 4 are only available in
electronic form at the CDS via anonymous ftp to cdsarc.u-strasbg.fr 
(130.79.128.5)
or via http://cdsweb.u-strasbg.fr/cgi-bin/qcat?J/A+A/}}

\author{A. L. Chies-Santos\inst{1}, M. G. Pastoriza\inst{1}, B. X.
Santiago\inst{1} \and Duncan A. Forbes\inst{2}} 

\offprints{ana.leonor@ufrgs.br} 

\institute{Departamento de
Astronomia, Instituto de F\'{\i}sica, UFRGS. Av. Bento Gon\c
calves 9500, Porto Alegre, RS, Brazil \\
\and Centre for Astrophysics \& Supercomputing, Swinburne University, Hawthorn VIC 3122, Australia}

\date{Received --; accepted --}

\abstract
{NGC\,5846 is a giant elliptical galaxy with a previously well
studied globular cluster system (GCS), known to have a bimodal
colour distribution with a remarkably high red fraction.
Here we revisit the
central galaxy regions searching for new globular cluster (GC)
candidates, and measure magnitudes, colours and sizes
for them. We also search for their X-ray counterparts.
We use archival Hubble Space Telescope WFPC2 images, from which 
we modelled and subtracted the host light distribution to increase 
the available sample of GCs. We performed photometry on the central objects, 
and measured sizes and coordinates for the entire GC 
system known in this galaxy.
We detect two dozen previously unknown GC candidates in the central regions.
Reliable sizes are obtained for about 60 GCs; their
typical effective radii are in the range $\rm 3-5$ pc.
The largest clusters are located in the central regions. 
We find 7 X-ray counterparts to GCs, most of them in
the central region. They are among the most luminous X-ray
sources in NGC\,5846. They are also optically luminous, compact
and belong to the red subpopulation.
The GCS of NGC 5846 is composed of relatively luminous X-ray sources.}

\titlerunning{The GCS of NGC\,5846 Revisited}

\authorrunning{Chies-Santos et al.}

\maketitle

\section{Introduction}

NGC\,5846 is a giant elliptical galaxy at the center of a large 
group of galaxies. It has a radial velocity of $\rm 1714 (\pm 5)
km\,s^{-1}$, a distance modulus of (m-M)=$\rm 32.32(\pm 0.23)$ and $\rm
M_{V}=-21.57$ (Forbes, Brodie \& Huchra (\cite{forbes96}). 
It is classified as a transition AGN  by Merrifield
(\cite{mer04}). The globular cluster system (GCS) of this galaxy
was studied by Forbes, Brodie \& Huchra (\cite{forbes96},
\cite{forbes97}) (FBH1 and FBH2 respectively, or simply FBH when
we refer to both papers at once). It was found that NGC 5846 has
a much lower specific frequency of globular clusters (GCs) than
other dominant ellipticals. On the other hand,
similarly to many other luminous early-type galaxies, its GCS has
a bimodal colour distribution. The blue and red peaks of the
distribution are located at V--I=0.96 and 1.17, respectively,
which roughly corresponds to a metalicity of [Fe/H]=--1.2 and
--0.2. This behaviour is often interpreted as the result of
several episodes of star formation or merging contributing to the GCS formation 
(e.g. Ashman \& Zepf \cite{ashman92}; Forbes,
Brodie \& Grillmair \cite{fbg97} and C\^ot\'e, Markze \& West {\cite{cote98}}). Interestingly, FBH found a very high
ratio of red to blue GCs. They  
also observed a smooth radial colour gradient in the sense
that the relative fraction of blue GCs increases outwards, something that
has also been observed in other galaxies. The central
region of NGC\,5846, however, is a difficult environment for
detecting GCs, even in Hubble Space Telescope (HST) images, given
the presence of dust and the high central optical surface
brightness. Contrary to other galaxies (eg. Larsen \textit{et al.} \cite{larsen01} and Jord\'an \textit{et al.} \cite{jordan05}), no GC size
determinations are so far available for NGC\,5846.

Chandra observations of NGC\,5846 reveal a complex X-ray
morphology of the hot gas.  These high resolution images confirm
the previously reported similarity between the distribution of
X-ray and the H$\rm\alpha$ + [NII] emission, indicating that hot and
warm gases are linked (Trinchieri \& Goodfrooij
\cite{trinchieri02}). The hard X-ray nucleus is coincident with
the radio core observed with the VLA, but is displaced relative
to the optical centre (Filho et al. \cite{filho04}). A population
of 41 individual X-ray sources was also observed by Trinchieri \&
Goodfrooij (\cite{trinchieri02}), with X-ray luminosities ($\rm L_{X}$) in
the range of $\rm3\times10^{38} - 2\times10^{39}
erg\,s^{-1}$. The luminosity distribution of these sources
appears steeper than in any other early type galaxy studied to date.
In our Galaxy about 10\% of the luminous X-ray sources are found
to be associated with globular clusters, which may harbour X-ray
binaries (Verbunt \& Hut \cite{verbunt87}). A large fraction
of the resolved X-ray point sources in early-type galaxies are likely
to be low mass X-ray binaries (LMXBs). Angelini, Loewenstein \& Mushotzky (\cite{angelini01})
found that about 70\% of LMXBs in NGC\,1399 have a spatial connection with GCs. 
For NGC\,4697 this fraction decreases to 20\% (Sarazin, Irwin, \& Bregman (\cite{sarazin01}).

The main goal of this paper is to revisit the central regions of
NGC\,5846 using archival HST images, searching for new GC candidates
which may have gone undetected, and investigating their optical
properties. We also aim to measure GC sizes and to find any possible
link between GCs and X-ray point sources detected by Trinchieri
\& Goodfrooij (\cite{trinchieri02}).  This paper is organized as
follows: In Sect. 2 we summarize the observations and basic data
reduction.  In Sect. 3 we present the photometry and discuss the
colour distribution of the central GCs. In Sect. 4 we determine
the sizes of the GCs, assessing the reliability of these
measurements. In Sect. 5 we compare the coordinates of
the X-ray point sources and the GCs. We present our summary and
concluding remarks in Sect. 6.

\section{Observations and data reduction}
Archival HST V and I images from program G0-5920 were used to
identify and study in detail GC candidates. The images were
obtained with the Wide Field Planetary Camera 2 (WFPC2) in the F555W
and F814W filters. There are three different pointings,
corresponding to 4 images in each pointing, 2 for each filter,
resulting in a total of 12 images, see Table.~\ref{jobs}

\begin{table}
\centering
\renewcommand{\tabcolsep}{1.0mm}
\begin{tabular}{c c c c c}

\hline\hline
Rootname & Filter & Pointing & Exposure Time (s)\\
\hline
U36J0401T & F555W & Central & 900\\
U36J0402T & F555W &         & 1300\\
U36J0403T & F814W &         & 900\\
U36J0404T & F814W &         & 1400\\
U36J0501T & F555W & North   & 900\\
U36J0502T & F555W &         & 1300\\
U36J0503T & F814W &         & 900\\
U36J0504T & F814W &         & 1400\\
U36J0601T & F555W & South   & 900\\
U36J0602T & F555W &         & 1300\\
U36J0603T & F814W &         & 900\\
U36J0604T & F814W &         & 1400\\
\hline
\end{tabular}
\caption{Journal of Observations - HST/WFPC2 images}
\label{jobs}
\centering
\end{table}

The spatial scale is $\rm0.046 \arcsec\,pixel^{-1}$ and the field is of
37\arcsec x 37\arcsec~for the planetary camara (PC). For the Wide Field Camera (WFC)
the scale is $\rm0.1\arcsec\,pixel^{-1}$, corresponding to a 80\arcsec
x 80\arcsec~field. The images listed in Table~\ref{jobs} had
been previously reduced and used to study the GCS by
FBH. These authors measured CCD positions, V, I magnitudes and
colours for a sample of a little less than 1000 GC.
Our main goal in this
work was to investigate in more detail the GC properties in the
inner regions of NGC\,5846. We therefore carried out a careful
analysis of the PC images of the central pointing. The initial
data reduction process was similar to that of FBH1. We aligned
the pairs of individual exposures with each filter and then
combined then using the STSDAS task \textit{gcombine}. In order
to identify the GCs in the central region of the galaxy, we
carried out detailed surface photometry of the host galaxy. We
first built a model of the luminosity distribution of the galaxy
using the STSDAS task \textit{ellipse} for each filter. We then
subtracted these models from the V and I PC images that resulted
from \textit{gcombine}. The results of the subtraction reveal a
very conspicuous filamentary dust morphology as well as numerous
bright compact objects. The sample selection was made by eye
using the V subtracted image. 

\begin{figure}
\resizebox{\hsize}{!}{\includegraphics{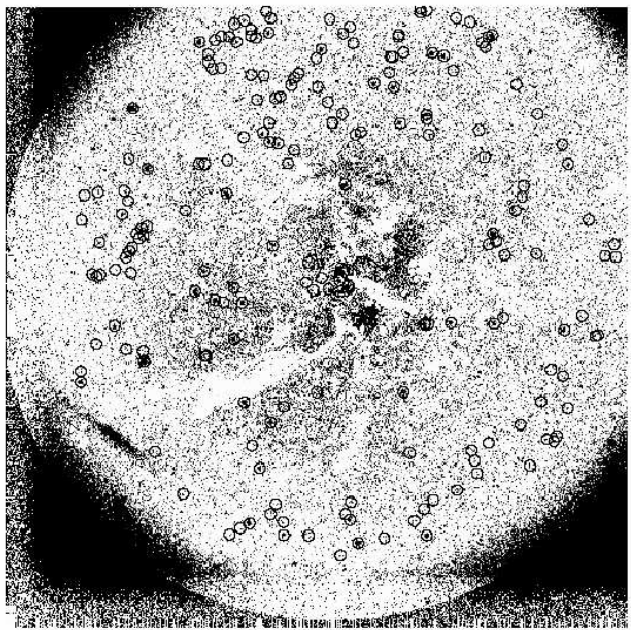}}
\caption[]{The result of the subtraction of
the model from the \textit{gcombined} V image in grey scale, 37\arcsec on a side. The eye selected GCs canditades are indicated by circles. North is at about $70^\circ$ from the vertical going counterclockwise}
\label{sub}
\end{figure}

Figure
\ref{sub} shows the model subtracted image in the V-band. The
visually selected objects are marked on the image. We selected a
total of 196 sources on the subtracted image, 31 of which are in
common with FBH. Despite our careful selection, this number is
likely to be only a fraction of the total, since it is very
complex to determine the exact number of extragalactic globular
clusters because of the dificulty to detect them against the
bright background of the central regions of their host
galaxies.

From the subtracted image we notice that the dust is strongly
concentrated in the very center of the galaxy, having an
assymetric distribution. In this dusty part of the galaxy the
stellar objects appear completely absent.  FBH2 suggested that
the dust is the result of a past merger or accretion of a
gas-rich galaxy and provides the fuel to the compact radio core,
which can be seen in both V and I images.

\section{Photometry}

\begin{figure}
\resizebox{\hsize}{!}{\includegraphics{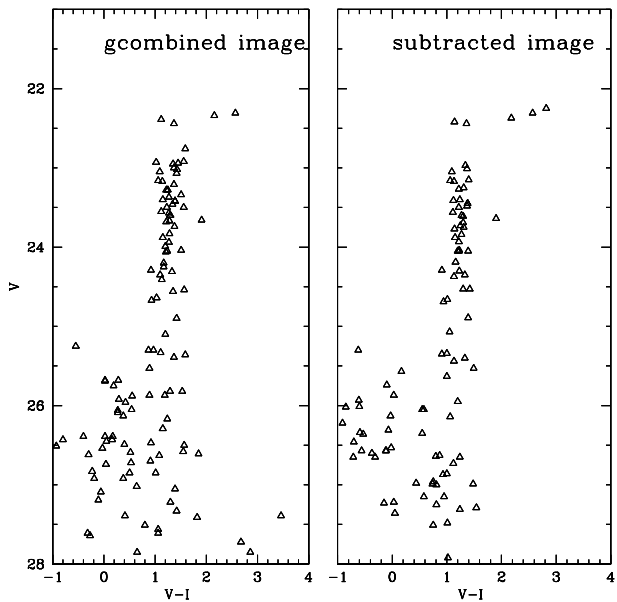}}
\caption[]{Left panel: CMD for the \textit{gcombined} image; right panel: CMD for the subtracted image.}
\label{cmd}
\end{figure}

Having CCD positions we then measured magnitudes using the APPHOT
task \textit{phot}, in both \textit{gcombined} images and the
galaxy subtracted ones. In Figure \ref{cmd} we show the
$\rm V, I$ colour
magnitude diagrams (CMD) for all visually selected compact
objects obtained from both images. Both CMDs exibit a well defined
locus of GCs spanning the range $\rm 22.5< V <27.5 $ and $\rm
0.5<V-I<1.9$.  The mean V--I colour of $\sim$1 is
typical of GCs. A 
total of 68 GC candidates are within these CMD limits. This GC
locus shifts blueward at fainter magnitudes as a selection
effect, since the sample was visually selected in the V
image. Also, the colour dispersion increases substantially for
$\rm V > 26.5$ due to photometric errors. Notice that the dispersion in colours tends
to be smaller in the subtracted images, attesting to the higher
photometric precision achieved when subtracted images are
used. We therefore used the subtracted images for our photometry
and cut the GC sample at $\rm V = 26.5$.

In order to have photometric measurements as consistent as
possible with those from FBH1, our photometry was carried out in a
similar way; we adopted a 2 pixel aperture radius, the aperture
corrections to a 0.5\arcsec radius were of 0.39 mag (for F555W)
and 0.54 mag (F814W). These values were based on Table 2 of
Holtzman \textit{et al.} (\cite{holtz95a}). To convert to the
standard Johnson-Cousins V, I system for a gain ratio of 7 we
followed Table 7 of Holtzman\textit{et al.} (\cite{holtz95b}). We
finally corrected for Galactic extinction: $\rm A_V=0.11$ and
$\rm A_I=0.05$ (Faber \textit{et al.} \cite{faber89}), in order to be consistent with FBH previous work (the Schlegel, Finkbeiner \& Davis (\cite{sfd98}) values are $\rm A_V=0.182$ and
$\rm A_I=0.107$).

\begin{figure}
\resizebox{\hsize}{!}{\includegraphics{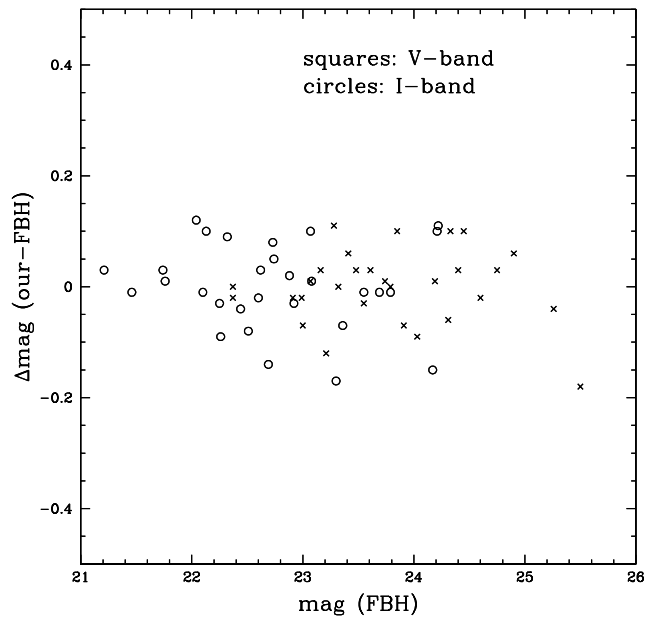}}
\caption[]{Photometric differences after comparison between our photometry and that of FBH, after correcting for the fitted linear relation (see text).}
\label{compforb}
\end{figure}

Since our final goal is to combine our inner GC sample with the
larger sample by FBH we compared our magnitudes in both filters
for the common objects. A small offset of $\rm \simeq 0.1 mag$
was found for both filters.
Our final calibrated magnitudes were
converted to the FBH1 photometric system by fitting a linear
relation of the type $\rm m_{FBH1}=am_{our} + b$, where $\rm
m_{FBH1}$ is the magnitude as measured by FBH1 and $\rm m_{our}$
is our corresponding measurement. For the V-band coefficients we
obtained $\rm a=0.984$ and $\rm b=0.287$; for the I-band
coefficients we found $\rm a=0.947$ and $\rm b=1.232$. The fits
were carried out only for GC candidates with quality photometry
($\rm V<26.5$). In Figure \ref{compforb} we show the residual
differences as a function of magnitude for both filters. The scatter of the residuals are $\rm \pm0.07$ for the V-band and $\rm \pm0.08$ for the I-band.

\subsection{Central Colour Distribution}

Colour distributions are very useful for characterizing GC
populations. They have been observed to display a bimodal pattern
in large early-type galaxies either in clusters or in the field
(Elson \& Santiago \cite{elson96} and Rhode \& Zepf
\cite{rhode04}).
NGC\,5846 also presents this behaviour, as already pointed
out by FBH. These authors also find a variation in the GC colour
distribution as a function of galactocentric distance, which they
interpret as a metallicity gradient, in the sense that metallicity
decreases outwards. Here we take advantage of the increased data
sample in the inner regions of the same galaxy and review the GC
colour distribution.

\begin{figure}
\resizebox{\hsize}{!}{\includegraphics{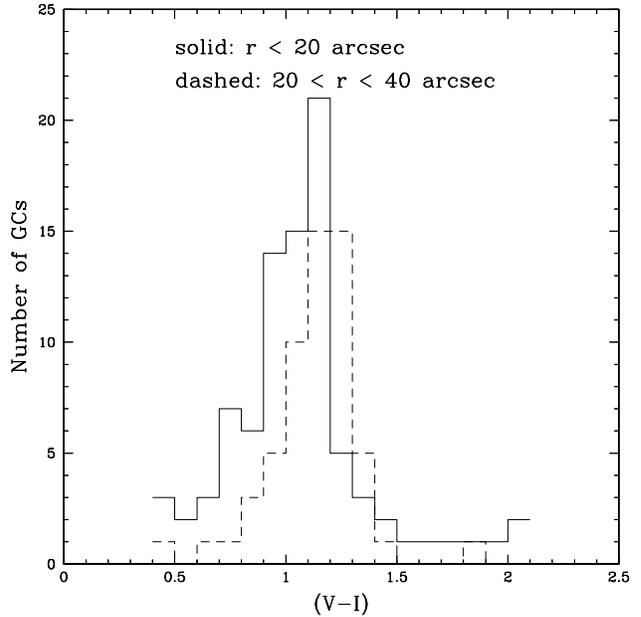}}
\caption[]{Colour distributions at the central parts of NGC\,5846. Solid line: $\rm r<20\arcsec$; dashed line: $\rm 20\arcsec < r < 40\arcsec$ }
\label{coldist}
\end{figure}

Figure \ref{coldist} shows the colour distribution of GCs with
$\rm V<26.5$ for two different rings from the center. Both colour
distributions shown are dominated by typical GC colours, ranging
from $ \rm 0.6 \leq (V-I) \leq 1.5$, showing that the GC sample
is not strongly contaminated. The inner population is shifted
slightly towards bluer colours relative to the one located further out
from the center of the galaxy.

One might think that dust could be making the GCs in the outer regions redder. Even though there is  dust 
in this region, as revealed by the lighter spots in Figure \ref{sub} (since its colour is inverted), these spots are clearly more 
pronounced close to the centre, and therefore should be affecting the inner 
($\rm r < 20\arcsec$) GCs more than the ones in the outer regions ($\rm 20\arcsec < r < 40\arcsec$) . Thus, correction for dust 
would actually increase the colour difference between the two regions being 
compared.

\begin{figure}
\resizebox{\hsize}{!}{\includegraphics {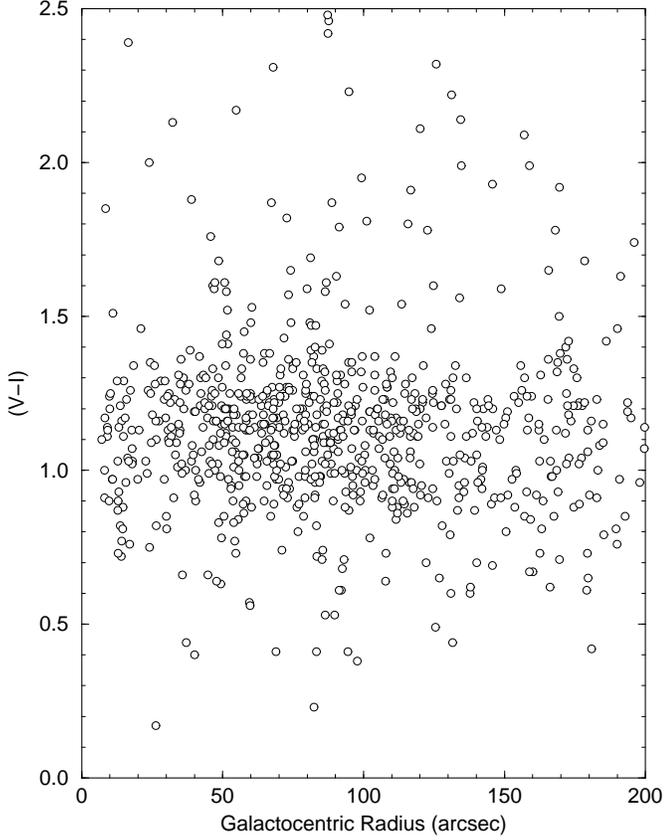}}
\caption[]{(V-I) colour of GCs vs Galactocentric radius}
\label{colourdistcent}
\end{figure}

The inner regions have a mean (V--I)
= 1.17 $\pm$ 0.04 compared to (V--I) = 1.24 $\pm$ 0.07 in the next
radial bin. Thus the difference ($\delta$(V--I) = 0.07) is small enough to place both regions within
the metal-rich GC subpopulation and it is 
of only marginal statistical
significance. We note that the original
data of FBH also had a hint for bluer colours at the very center of
NGC\,5846 (r $<$ 2 kpc), as indicated in their Figure 5.
If real it could be due to
a combination of age and/or metallicity effects. The inner
GCs could be slightly younger or slightly more metal-poor. It is
difficult to test either possibility without high signal-to-noise
spectra of these inner GCs, although we may use an SSP model (e.g. Bruzual \& Charlot \cite{bc03}) to
assess age and metallicity effects. Assuming an old population of 12 Gyrs
and [Fe/H] = --0.4 (i.e. similiar to that expected for red GCs), we can
investigate the age/metallicity differences needed to reproduce a
$\delta$(V--I) = 0.07 bluer colour. If the colour difference is purely due
to age, then the inner GCs are about 5 Gyrs younger. If purely due to
metallicity, the inner GCs would be more metal-poor by about 0.25 dex.
Such large differences would be easily detectable with high S/N spectra.
In practice, the bluer colour is likely to be a combination of age and
metallicity differences.
In Figure \ref{colourdistcent} we show a plot of (V-I) as a function of
galactocentric radius.

In Table 3 we list the 
positions and photometry for the entire GCS of NGC\,5846.

\section{GCs Sizes}

The determination of extragalactic GC sizes is extremely
difficult, even in HST data, given their extremely small sizes
compared to the extragalactic distances. 
Larsen \& Brodie (\cite{larsen03}) find that the blue
GCs are, on average, larger than the red ones. Red GCs are
generally smaller than blue GCs by about 20\% (Larsen et
al. \cite{larsen01}). In the central parts of galaxies this
difference is well pronounced. The effect may, at least in part,
arise from a GC size against 3D distance relation, combined with
different radial distributions between blue and red
GCs (Larsen \& Brodie (\cite{larsen03}). 
Studies of GC sizes are so far largely concentrated on
nearby galaxies ($\rm m-M\simeq31.5$ or less). In this section we
attempt to derive sizes for the NGC\,5846 ($\rm m-M\simeq32.32$) GCs.

\begin{figure}
\resizebox{\hsize}{!}{\includegraphics{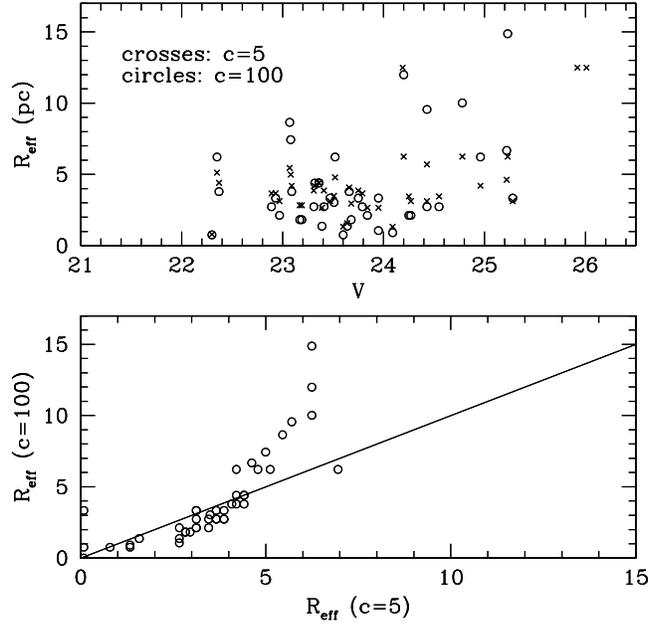}}
\caption[]{Upper panel: intrinsic size vs V-band magnitude relation; Lower panel: relation between effective radii as estimated from two extreme King models.}
\label{c5vsc100}
\end{figure}

We use the ISHAPE code described in Larsen
(\cite{larsen99}). This code convolves the WFPC2/HST point spread
function (PSF) with model King profiles in 2 dimensions. The
resulting model image is then compared to the real GCs and a $\rm
\chi^{2}$ is derived from the fit, along with an estimate of the
GC's intrinsic size. We have used an image taken at about
the same time as those listed in Table 1 to construct the PSFs
for the PC and WFC. We used DAOPHOT.PSF and DAOPHOT.SEEPSF tasks
for this purpose. We processed both PC and WFC objects, despite
the lower resolution and stronger undersampling of the latter. We
tested the sensitivity of the intrinsic sizes to the model
concentration parameter by running ISHAPE for the PC sample both
with c=5 and c=100, which are the extreme values available in the
code. In Figure \ref{c5vsc100} we show the effective radii ($\rm
R_{eff}$) in parsecs as a function of V magnitude for both $\rm
c$ values. We also compare the resulting $\rm R_{eff}$ values. We
see that there is no strong trend between size and magnitudes
regardless of the King model chosen. However, a slight trend in
the sense that brighter GCs are also larger ones is seen
brighter than $V \simeq 24$. Also notice that the $\rm R_{eff}$
values correlate with one another, except at large radii where
the c=5 models start to saturate. We therefore adopt the c=100
model in this paper.

\begin{figure}
\resizebox{\hsize}{!}{\includegraphics{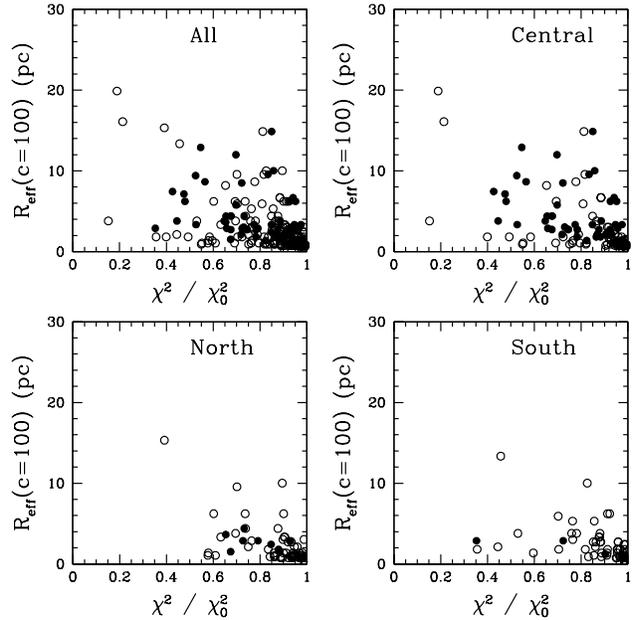}}
\caption[]{Size as a function of $\rm \chi^{2}/\chi_{0}^{2}$. Solid symbols represent objects located in the PC chips, whereas open symbols are for WFC chips. Upper left: the entire sample; Upper right: central pointing; Lower left: north pointing; Lower right: south pointing. }
\label{chiratio}
\end{figure}

In Figure \ref{chiratio} we plot the sizes of GCs as a function
of $\rm \chi^{2}/\chi_{0}^{2}$, where $\rm \chi_{0}^{2}$ is the
result of fitting a point source (delta function) rather than a
King profile to the objects. Thus, this ratio quantifies the
improvement obtained with the King model relative to a unresolved
source. All objects in the sample from the three pointings are
shown in the different panels. We note that the larger the $\rm
\chi^{2}/\chi_{0}^{2}$ the smaller becomes the radius, as an
effect of not resolving the GC from a point source. In
particular, objects with $\rm \chi^{2}/\chi_{0}^{2} > 0.8$ are
essentially unresolved as attested by their systematically small
$\rm R_{eff}$. We therefore cut the sample at $\rm
\chi^{2}/\chi_{0}^{2} \leq 0.8$.

\begin{figure}
\resizebox{\hsize}{!}{\includegraphics{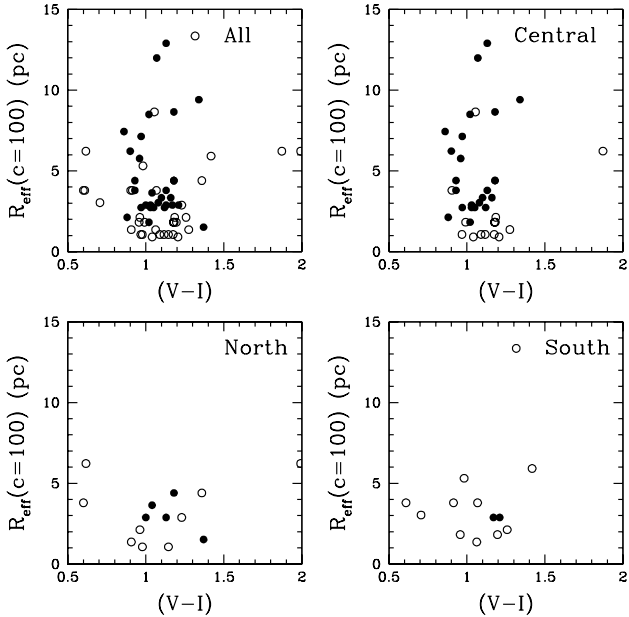}}
\caption[]{Size as a function of colour for all 3 individual
pointings and for the combined pointing. Open circles are PC GCs and closed ones are WFC GCs.} 
\label{colreff}
\end{figure}

Next we explore the possibility of a size--colour relation
similar to that found by previous authors. In Figure
\ref{colreff} we plot the sizes of GCs, cut at $\rm
\chi^{2}/\chi_{0}^{2} < 0.8$, as a function of (V-I) colour,
again separating the different pointings in different
panels. Even though a relatively small number of GCs is
available (especially blue GCs), given our selection criteria, there is no obvious
trend in our data for a relation between size and colour. Notice,
however, that the larger GCs are seen in the central pointing,
specially at the central PC: only a couple of GCs with $\rm
R_{eff} > 5pc $ are seen in the South or North pointings, which
are located $\sim 2\arcmin $ from the center.

Finally, the typical size of our resolved GCs is between 3 and 5
pc, with only a handful being larger than 10 pc. These values
are typically larger than the median GC sizes found by Larsen et
al. (\cite{larsen01}) in their study of nearby galaxies. The
difference is likely caused by selection effects, given the
larger distance modulus of NGC\,5846. We note that there is still
some debate in the literature as to whether such size differences
are due to projection effects (e.g. Larsen \& Brodie
(\cite{larsen03}; Forbes et al. 2005; Spitzer \textit{et al.} 2006) 
or mass segregation (Jordan 2004). 

Table 4 lists the sizes and related parameters for those GCs
which were successfully measured by ISHAPE.

\section{Comparing the GCs and the X-Ray Point Sources}

More than thirty luminous X-ray sources associated with Galactic
globular clusters have been observed with Einstein, ROSAT,
XMM-Newton and Chandra, since 1975 and up to this date (Verbunt
\& Lewin \cite{verbunt04}). On the other hand, high resolution
observations made by ROSAT and Chandra in elliptical and spiral
galaxies, such as NGC\,4697, NGC\,4472, M31 and NGC\,4594, reveal
X-ray sources associated with GCs.  Irwin, Bregman \& Athey
(\cite{irwin04}) classify ultraluminous X-ray sources (ULXs) as
having $\rm L_{X} = 1 - 2 \times 10^{39}erg\,s^{-1}$ and very
ultraluminous X-ray sources (VULXs) as having $\rm L_{X} > 2
\times 10^{39}erg\,s^{-1}$. In the early type galaxy NGC\,1399
more than two hundred sources were resolved in a
8\arcsec~x~8\arcsec~region, of which 45 were found to be
associated with globular clusters and two were found to have $\rm
L_{X} > 2 \times 10^{39}erg\,s^{-1}$ (Angelini, Loewenstein \&
Mushotzky \cite{angelini01}), therefore being VULXs.  In
NGC\,5846 forty one individual X-ray sources were observed with
Chandra X-ray Observatory (Trinchieri \& Goodfrooij
\cite{trinchieri02}). All of them have $\rm L_{X} > 3 \times 10^{38}
erg\,s^{-1}$. These $\rm L_{X}$ values are above or close to the
Eddington luminosity for a 1.4 solar mass accreting object
(Sarazin et al. \cite{sarazin03}), suggesting that, if caused by
a single X-ray binary, they are accreting onto a black hole,
rather than a neutron star. On the other hand, in the case of a
globular cluster, more than one LMXBs may be present, mimicking
the effect of VULXs.

Here we test the hypothesis that some of these point sources
found by Trinchieri \& Goodfrooij (\cite{trinchieri02}) (Table 1
of their paper), are associated with globular clusters in the
field of view of HST/WFPC2 studied here. Equatorial coordinates
(J2000) were obtained from the WFPC2 images with the STSDAS
\textit{xy2rd} IRAF task. This astrometric solution, however, may
be systematically offset from the Chandra positions by some
unknown amount. Therefore we searched for X-ray counterparts of
our GCs at the same time as we found the best estimate for these
offsets. We did this by finding the offsets in RA and DEC that
maximized the number of matches between optical and X-ray
sources. A total of 7 such matches resulted. The corresponding
offsets were of $\rm 0.8\arcsec$ in RA and $\rm 0.2\arcsec$ in
DEC. These values are consistent with previous astrometric
comparisons between HST/WFC2 and Chandra (Jord\'an et
al. \cite{jordan04}; Xu et al. \cite{xu05}).

\begin{figure}
\resizebox{\hsize}{!}{\includegraphics{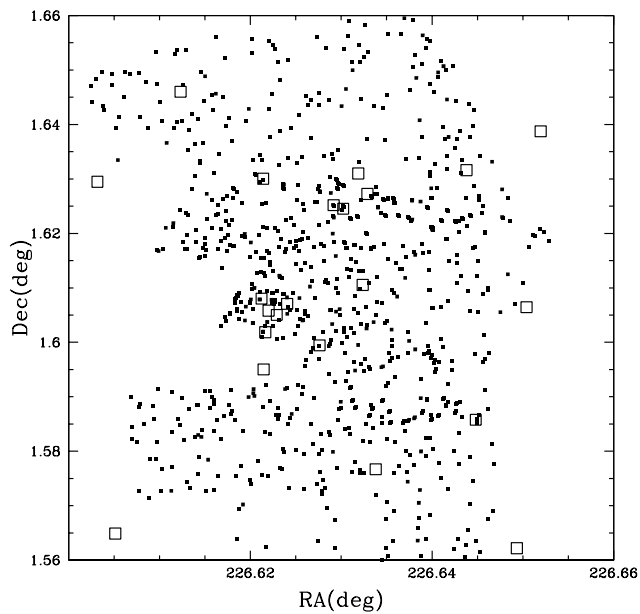}}
\caption[]{On-sky map of X-ray sources and GCs in the direction of NGC\,5846. The small points represent the GCs and the large squares correspond to the Chandra point sources.}
\label{matox}
\end{figure}

We have found GC/X-Ray sources in all three pointings, although
most matches are located in the central WFPC2 field. In
Fig. \ref{matox} we show the on-sky distribution of both optical
and X-ray objects, with positions already compensated for the
offsets mentioned above. We find that the matched objects tend to
concentrate to the inner parts of the galaxy. In Table \ref{xgc}
we list the main properties of these GC/X-ray objects. The listed
positions are those from Trinchieri \& Goodfrooij
(\cite{trinchieri02}). Their colours show that they belong to the
red population, with no exception. This result is in agreement to
what Angelini, Loewenstein \& Mushotzky (\cite{angelini01}) found
for NGC\,1399. We also notice that the V magnitudes of 5 matched
objects are brighter than 23.6, which means that the X-ray
emitting GCs are usually brighter than average, as previously
found by Sarazin et al.(\cite{sarazin03}) and Kundu et
al.(\cite{kundu02}) for other luminous ellipticals.  Considering
their intrinsic sizes, only three of the matched objects were
resolved with the ISHAPE code (Sect.4), but 2 of them were
marginally resolved; all three are smaller than $\rm R_{eff}=2.8
pc$, being therefore bright but compact GCs.

In terms of X-ray properties, several of our matched GCs lie among the most
luminous objects in the host galaxy. Source 16 is a ULX and sources 17 and 24 have measured $\rm L_{x}$ close to $\rm 10^{39} erg
s^{-1}$, being among the 25\% brightest X-ray sources in
NGC\,5846. They are therefore likely to be either a single binary
system accreting onto a black hole or several LMXBs at or near
the Eddington luminosity, as discussed in Angelini, Loewenstein
\& Mushotzky (\cite{angelini01}). Besides, the X-ray sources 8 and 11 have 
been classified by Trinchieri \& Gooudfrooij (\cite{trinchieri02}) as extended regions. Alternatively, 
they could be composed 
of several overlaping sources having different origins. Because they have total luminosity of $\rm L_{x} >  
2\times 10^{39} erg\,s^{-1}$, the corresponding GC match may still be a ULX.

We note that NGC\,5846 has been described as a complex X-ray
object by Filho et al. (\cite{filho04}). They find a coincidence
between the VLA radio core and the X-ray nucleus of the galaxy;
this latter position is where source number 12, according 
to classification of Trinchieri \& Goodfrooij
(\cite{trinchieri02}), is located. Notice that this is one among
several other luminous X-ray sources. Also, we observe that the
position of source 12 is not coincident with the optical center.

Since VULXs are found to be associated with late rather than
early-type galaxies, NGC\,5846 seems to be an exception along
with NGC\,1399.

\begin{table}
\centering
\renewcommand{\tabcolsep}{1.0mm}
\begin{tabular}{c c c c c c }
\hline\hline
 Source & $\alpha$ &$\delta$ & $\rm V $& $\rm(V-I)$& $\rm Lx$ \\
	& (J2000) & (J2000) & (mag) & (mag) & $\rm (erg\,s^{-1})$\\
\hline
8  &15:06:29.15& +01:36:28.7& 22.8 & 1.12  & $\rm 9.32 \times 10^{39}$\\
9  &15:06:29.19& +01:37:47.9& 25.23& 2.42&$\rm 4.2 \times 10^{38}$\\
11 &15:06:29.25& +01:36:06.5& 23.39&1.17  &$\rm 1.77 \times 10^{39}$\\
15 &15:06:30.68& +01:35:57.8& 23.62& 1.29&$\rm 4.2 \times 10^{38}$ \\
16 &15:06:31.05& +01:37:30.4& 23.23& 1.12&$\rm 1.22 \times 10^{39}$\\
17 &15:06:31.31& +01:37:28.1& 24.32&1.26 &$\rm 6.5 \times 10^{38}$ \\
24 &15:06:34.8&  +01:35:08.6& 23.16&1.15 &$\rm 7.8 \times 10^{38}$ \\
\hline
\end{tabular}
\caption{The globular clusters coincident with Chandra X-ray sources }
\label{xgc}
\centering
\end{table}

\section{Summary and Concluding Remarks}

With V and I WFPC2 images taken from the Hubble Space Telescope public
archive we have detected 68 central globular clusters within
the PC ($\rm r\leq 2kpc$); 23 of these are previously
undetected objects. We combined our inner sample with the larger
data set by FBH and revisited the properties of this central
sample. We have determined the coordinates and
photometry for all GCs
in the combined sample and also attempted to measure their
intrinsic sizes. For several dozens, useful estimates of the
effective radius were obtained, especially in the PC image. 
This information is given in two electronic tables. 
We also searched for X-ray counterparts to the
GCs in the Chandra sample of Trinchieri \& Goodfrooij
(\cite{trinchieri02}). 

Our main results are:

\begin{enumerate} 
\item The colour distribution shows a hint of becoming bluer
in the very central galaxy regions.
\item Most GCs with reliable size estimates have  $\rm R_{eff} >
3$ pc; larger GCs tend to be seen in the central region of the
host galaxy. In fact, only a few GCs with $\rm R_{eff} > 5$ pc
are located outside the central pointing. This is opposite
to the behaviour found by van den Bergh, Morbey, \& Pazder
(\cite{van91}) for the Milky Way GCs. We note however, that all
three early-type galaxies with extensive spatial coverage in the
sample of Larsen et al. (\cite{larsen01}) show an increase in
median GC $\rm R_{eff}$ in the most central radial bin, especially
for the blue sub-population. The central pointing also has the largest total number of GCs, so if the 
large GCs are just the tail of the GC size distribution, then one would 
expect an statistical excess of these large GCs where there are more GCs 
in general. However, the excess of large GCs in the central pointing
remains after we correct for this effect:
the outer fields have about 0.4 times the number of GCS in the central pointing, but they have almost no GCs with $\rm R_{eff} > 5 pc$.
\item No clear evidence for a size-colour relation is found.
This result does not seem to agree with the findings of Larsen et
al. (\cite{larsen01}), who observed that the blue clusters are
generally larger than the red ones by about 20\%. However 
this is possibly a selection effect as NGC 5846 is more distant their 
more nearby galaxies (for which smaller intrinsic
sizes could be measured) and our
sample has very few blue GCs. 
\item From a positional match between optical and X-ray coordinates, we found 7 GC/X-ray matches, most of which are located in the central parts. Optically, they tend to be bright ($\rm V\leq 23.5$) and compact, since only one was clearly resolved. All of them have $\rm (V-I) > 1.1$, which places them as members of the red sub-population. Their X-ray luminosities are also among the highest in the Chandra sample, with 3 of them being among the 25\% most luminous and 2 other matches lying in complex central regions with $\rm L_{X} > 10^{39}erg\,s^{-1}$.

\end{enumerate}

The X-ray emitting GCs are typically more luminous in X-rays than
those found in other early-type galaxies. For instance Angelini,
Loewenstein \& Mushotzky (\cite{angelini01}) found that 70\% of
the X-ray point sources in the direction of NGC\,1399 were
located within GCs, but almost all of them have $\rm L_{X} <
10^{39} erg\,s^{-1}$. Our lower detection rate of GC/X-rays in
comparison to theirs, associated with a higher fraction of luminous sources,
suggests that the GCS of NGC\,5846 is composed
of relatively few, although very luminous X-ray sources. 

This
conclusion also seems to hold when NGC\,5846 is compared to other
ellipticals, either very luminous ones, such as NGC\,4472, or
more typical, such as NGC\,4697 (Kundu, Macarrone \& Zepf
\cite{kundu02}; Sarazin, Irwin \& Bregman
\cite{sarazin00}).
At the high end of the
$L_{X}$ luminosity function, Trinchieri \& Goodfrooij
(\cite{trinchieri02}) found 3 ULXs and 4 extended regions,
with high total $\rm L_{X}$ out of 41 X-ray point like sources, where
around 2 would be expected to be foreground/background sources
(Irwin, Bregman \& Athey \cite{irwin04}). NGC\,5846 displays an
extended and complex structure in X-ray emission, described by
Filho et al. (\cite{filho04}), something that is also observed in
other early-types, like NGC\,1600 (Sivakoff, Sarazin \& Carlin
\cite{siva04}). We are currently analysing images obtained from
the New Technology Telescope (NTT) and HST archive of NGC\,1600
in order to check if there is any similarity to NGC\,5846. A
correlation between the overall galactic X-ray properties and the
GCS would provide useful additional constraints to the processes
that govern the formation of both GCs and their host galaxy.

\begin{acknowledgements}
We acknowlodge the financial support of CNPq and the comments and
suggestions of an annonymous referee. DF thanks the ARC for financial support. We thank J. Irwin for
comments and suggestions that helped improve the paper.

\end{acknowledgements}

\end{document}